\documentclass[lettersize,journal]{IEEEtran}
\usepackage{amsmath,amsfonts}
\usepackage{algorithmic}
\usepackage{algorithm}
\usepackage{array}
\usepackage[caption=false,font=normalsize,labelfont=sf,textfont=sf]{subfig}
\usepackage{textcomp}
\usepackage{stfloats}
\usepackage{url}
\usepackage{verbatim}
\usepackage{graphicx}
\usepackage{cite}
\usepackage{booktabs} 
\usepackage{url}
\usepackage{xspace}
\usepackage[font=normal]{caption}
\usepackage{subcaption}
\usepackage{float}
\usepackage{textcomp}
\usepackage{multirow,makecell}
\usepackage{pifont}
\usepackage{tikz}
\usepackage{bbm}
\usepackage[inline]{enumitem}
\usepackage{adjustbox}
\usepackage{xcolor}
\usepackage{listings}
\usepackage{xcolor}

\lstdefinestyle{codeSnippet}{
  numbers=left,
  backgroundcolor=\color{gray!5},
  numberstyle=\tiny\color{gray},
  xleftmargin=2em,
  stepnumber=1,
  numbersep=8pt,
  basicstyle=\ttfamily\small,
  keywordstyle=\color{blue},
  commentstyle=\color{green!50!black},
  stringstyle=\color{orange},
  frame=single,
  breaklines=true
}

\usepackage{minted}

\usepackage{tikz}
\usetikzlibrary{tikzmark}
\usetikzmarklibrary{listings}
\usepackage{amsmath}
\usepackage{amssymb}

\usepackage{amsmath}
\usepackage{graphicx}
\usepackage{makecell}

\usepackage{colortbl}
\definecolor{darkgreen}{rgb}{0.0, 0.3, 0.13}
\definecolor{darkred}{rgb}{0.2, 0.0, 0.13}
\usepackage{tikz}
\usepackage{tabularx}
\usepackage{colortbl}

\usepackage{orcidlink}
\usepackage[utf8]{inputenc}
\usepackage{textgreek}
\usepackage{textcomp}

\DeclareUnicodeCharacter{2265}{\ensuremath{\geq}}

\usepackage{pifont}
\usepackage{xspace}
\usepackage[T1]{fontenc}
\newcommand{\gpt}{{\tt gpt-5}\xspace}
\newcommand{\gemini}{{\tt gemini-2.5}\xspace}
\newcommand{\claude}{{\tt claude-4.5}\xspace}

\newtheorem{takeaway}{Takeaway}
\newcommand{\SonarRule}[1]{\texttt{#1} rule}

\newcommand{\etal}{{\em et al.}\xspace}

\newcommand{\BfPara}[1]{{\noindent\bf #1.}\xspace}
\newcommand{\eg}{{\em e.g.},\xspace}
\newcommand {\ie}{{\em i.e.},\xspace}

\newcommand{\ours}{MA-CoT\xspace}

\usepackage{hyperref}

\hypersetup{
	plainpages=false,
	colorlinks,
	urlcolor=blue,
	linkcolor=red,
	citecolor=green,
	bookmarksnumbered
}

\begin{document}

\title{Enhancing Reliability in LLM-Based Secure Code Generation}

\author{Mohammed~F.~Kharma, Mohammad~Alkhanafseh, Ahmed~Sabbah, David~Mohaisen\IEEEmembership{, Senior Member,~IEEE}
\thanks{M. Kharma, M. Alkhanafseh, A. Sabbah, and D. Mohaisen are with with the Department of Computer Science, Birzeit University, Ramallah, Palestine. D. Mohaisen is also with the Department of Computer Science, University of Central Florida, Orlando, FL 32816 USA.
E-mail: mohaisen@ucf.edu (Corresponding author)}}

\maketitle

\begin{abstract}
Large language models (LLMs) are widely used for code generation, but their security reliability remains inconsistent across languages and prompting strategies. Existing prompt engineering improves functional correctness but rarely ensures consistent security outcomes. We introduce the \textit{Mitigation-Aware Chain-of-Thought (MA-CoT)} framework, which embeds task-specific CWE mitigation guidance and language-aware safeguards to reduce recurring vulnerabilities in generated code.   We evaluate MA-CoT across three LLMs (gpt-5, claude-4.5, gemini-2.5), three programming languages (C, Java, Python), and four prompting strategies (Vanilla, Zero-shot, CoT, MA-CoT) on a 200-task primary dataset, with external validation on LLMSecEval. Using static analysis with expert validation, MA-CoT reduces total security findings from 92 to 39 (57.6\%) on the primary dataset and from 73 to 4 (94.5\%) on LLMSecEval. High-severity findings (Blocker + Critical) drop from 90 to 39 (56.7\%) and from 45 to 2 (95.6\%), respectively.  Across both datasets, MA-CoT is the only strategy that consistently improves security reliability; Zero-shot and CoT are less reliable and may increase vulnerability, especially in C. We further introduce a strict layered attribution of vulnerability drivers (language-core vs. stack layers) and show that residual risk concentrates in hardening-oriented patterns (e.g., OS- and toolchain-dependent), motivating secure-by-construction primitives alongside prompting. Code and data are available at \href{https://github.com/mohsystem/paper3.git}{GitHub}.  
\end{abstract}


\section{Introduction}
\IEEEPARstart{L}{arge} language models (LLMs) are transforming software development by translating plain-language specifications into functionally correct code~\cite{ChenTJYDHKEHBJB21,LinM25,AsareNA24,MessaoudMJMG23,MedeirosNC22,JaffalAM25}. Tools such as GitHub Copilot, OpenAI Codex, and Google Gemini are now used daily by over 50\% of developers, providing productivity gains of 30–50\% on routine programming tasks~\cite{gitclear25, YetistirenOAT23}.

These efficiency gains are accompanied by concerns regarding the \textit{reliability} of LLM-generated code. Empirical evidence indicates that code produced by LLMs frequently contains security vulnerabilities, which can undermine system dependability. For example, Pearce \etal found that Copilot generated insecure implementations in roughly 40\% of security-critical scenarios, including mismanaged buffer boundaries, SQL queries, and cryptographic routines~\cite{PearceATDK22}. Fu \etal observed similar vulnerability patterns in Python and C++, which persisted even as model size and training data increased~\cite{FuLTLSYC25}. Sandoval \etal further showed that developers often accept insecure suggestions without sufficient review, particularly under time pressure, highlighting risks to the \textit{reliability and robustness} of automated code generation~\cite{SandovalPNKGD23}.

The risks of LLM-generated code extend beyond isolated errors. Asare \etal demonstrated that current LLMs have notable limitations in enforcing context-dependent security constraints, often generating code that appears reliable in isolation but fails to satisfy system-level security requirements~\cite{AsareNA23}. Ullah \etal reported similar findings via benchmark tests, showing that even explicit prompts rarely enable models to consistently identify or reason about vulnerabilities~\cite{ullah2024llms}. These results indicate fundamental gaps in how LLMs represent and apply security knowledge during code generation, undermining the overall \textit{reliability} of generated software.

Prompt engineering offers a method to steer model behavior without retraining. Techniques such as zero-shot, few-shot prompting, and chain-of-thought (CoT) reasoning can improve functional correctness and logical consistency~\cite{WeiWSBIXCLZ24, KojimaGRMI22}. However, {\em security and reliability considerations are mostly incidental}. Broad instructions like ``write secure code'' are too unspecific to consistently improve reliability, and they provide limited guidance for mitigating concrete vulnerabilities~\cite{TonyFMDS25}.

Our previous work~\cite{KharmaSAHM26} introduced the Weaknesses-Aware Chain-of-Thought (WA-CoT) method, which mapped programming tasks to relevant Common Weakness Enumeration (CWE) entries and embedded them in prompt guidance to inform the LLM of potential security flows. While promising, a broader evaluation revealed several reliability-related limitations:  
\ding{172} {\em Information Overload.} Large CWE catalogs can increase, rather than reduce, failure occurrences~\cite{AnZZLGLXK24}.  
\ding{173} {\em Coarse-Grained Integration.} Listing CWE identifiers and descriptions without concrete remediation steps seldom translates into secure and reliable code.  
\ding{174} {\em Language Agnosticism.} Uniform treatment of all programming languages overlooks differences in type safety, memory management, and language-specific reliability risks.  
\ding{175} {\em Lack of Reasoning Support Without Clear Mitigation.} Models may recognize vulnerabilities in principle but fail to apply security logic during actual code synthesis, compromising dependable code generation.

Motivated by these observations, we design \ours{}, a prompt-based method that improves the \textit{reliability} of LLM-generated code while remaining lightweight enough for practical deployment. \ours{} embeds reliability-oriented reasoning directly into the code generation process through three components: mitigation-focused CWE integration, a baseline security ruleset, and language-aware prompting.

Rather than providing extensive vulnerability descriptions, \ours{} uses \textbf{mitigation-focused CWE integration} to extract concise, actionable countermeasures from a structured CWE mitigation knowledge base. For each programming task, relevant security domains (e.g., cryptography, authentication, input validation) are identified, and targeted safeguards are incorporated into the prompt. For example, cryptographic operations trigger recommendations such as AES-256-GCM with AEAD guarantees, Argon2id key derivation with at least 128-bit salt, and nonce generation using a cryptographically secure random number generator. This approach ensures \textit{consistent reliability across tasks} without introducing prompt overload.

\ours{} further includes a \textbf{baseline security ruleset}, a universal set of secure coding principles applied across all tasks regardless of language. These include strict input validation, robust error handling, safe memory management, and defensive programming. Together, they provide a foundational reliability layer that reinforces dependable code generation independently of prompt specifics.

Finally, \ours{} employs a \textbf{language-aware prompting strategy}. Vulnerabilities manifest differently across programming languages, so the system augments a language-agnostic checklist with language-specific safeguards. In C, emphasis is placed on pointer initialization, bounds checking, and explicit memory clearing; in Python, the focus is input sanitization, safe deserialization, and proper exception handling. This alignment improves {\em reliability across diverse language contexts}.

A common source of confusion in language-aware secure generation is defining what constitutes a ``language feature.'' Many weaknesses arise not from the core language specification but from the surrounding stack, including standard libraries, third-party APIs, OS interfaces, and toolchains. To avoid over-attribution and ensure actionable guidance, we adopt a strict layered framework that separates language-core mechanisms from language-adjacent drivers, allowing us to identify and mitigate residual failure modes effectively.

\BfPara{Research Questions}
We evaluated the \ours{} using the following research questions:
\begin{description}
    \item[RQ1:] To what extent does targeted CWE-based mitigation reduce the severity of vulnerabilities in LLM-generated code compared to standard prompting techniques?
    
    \item[RQ2:] How do programming language characteristics influence the effectiveness of mitigation-aware prompting?
    
    \item[RQ3:] What vulnerability patterns emerge across models, and how can they inform model-specific security?
\end{description}

\BfPara{Contributions} The contributions of this study are as follows:

\begin{enumerate}
    \item \emph{\ours{} Prompt Method.} A prompting method that integrates actionable CWE mitigation guidance, a baseline security ruleset, and language-specific adaptations.
    
    \item \emph{Comprehensive Empirical Evaluation.} An extensive study across three state-of-the-art LLMs, three programming languages, four prompting strategies, and 200 security-relevant tasks, yielding 7{,}200 generated programs. Effectiveness is validated using LLMSecEval, a widely adopted benchmark for code generation~\cite{TonyFMDS25,DaiXG25,CotroneoLL25,LinWQCM25}.
    
    \item \emph{Security Metrics and Analysis.} A detailed assessment of vulnerability density, severity distribution, and CWE category across models, languages, and prompting strategies.
    
    \item \emph{Open Benchmark and Reproducibility.} A publicly released benchmark including task specifications, CWE mappings, and experimental artifacts to support replication and future research on secure code generation.
    
    \item \emph{Layered Language-Aware Attribution and Hardening Guidance.} A conservative attribution framework that distinguishes language-core, runtime, ecosystem, OS, and toolchain factors, providing actionable, layer-specific security strengthening recommendations.
\end{enumerate}

\BfPara{Organization} Section~\ref{sec:related} reviews the related work followed by our methodology in~\autoref{sec:method}, the dataset in~\autoref{sec:data}, the results in~\autoref{sec:results}, discussion in~\autoref{sec:discussion}, threat to validity in~\autoref{sec:validity}, and concluding remarks and future work in~\autoref{sec:conclusion}.

\section{Related Work}\label{sec:related}

LLM-based code generation is now used in end-to-end development workflow. However, security outcomes remain inconsistent across languages and prompt designs~\cite{PearceATDK22,LinWQCM25,HuynhL25, KlemmerHPLBPMRV24,TihanyiBFJC25}. Prior work~\cite{TonyFMDS25,SahooSSJMC24,ChenZLZ24,BasicG24,SchaadGB25,DattaAD25,KimBM23,LiRYY25,ShiZ25,ZhaoSHLGZL25,NazzalKKP24} has shown that prompting can steer functional behavior, yet security improvements are often inconsistent, model-dependent, and rarely driven by structured, task-specific mitigation knowledge.

Prompt engineering has evolved into a systematic discipline for shaping model behavior without parameter updates. Sahoo \etal~\cite{SahooSSJMC24} surveyed 41 prompt engineering techniques and highlighted the need for more application-centric prompt designs and combined prompting approaches beyond general-purpose ones. Chen \etal~\cite{ChenZLZ24} systematized prompting methods and described how structural prompt changes (\eg decomposition and multi-step prompting programs) can alter model behavior. In the context of secure coding, another review of the literature~\cite{BasicG24} emphasizes that security flaws in LLM-generated code cover a wide range of vulnerability types and cannot be fully mitigated by generic warnings alone. These works motivate security-specific prompt structures that inject actionable guidance rather than relying on broad reminders.

A major line of research treats LLMs as both generators and reviewers. Tony \etal~\cite{TonyFMDS25} provided a systematic investigation of prompting techniques for secure code generation and reported that Recursive Criticism and Improvement (RCI) can reduce weakness density (\eg lowered
the average weakness by 77.55\%  compared with GPT-4), while also noting that some weakness families (e.g., command injection patterns) may remain persistent under prompting. Their analysis showed that reductions are more reliable for certain CWEs (\eg CWE-94, CWE-259, and CWE-330) than for others. More broadly, the RCI paradigm was introduced as a general strategy for improving LLM task performance via iterative critique and revision~\cite{KimBM23}. Datta \etal~\cite{DattaAD25} studied reflection-style secure code generation at scale and quantified both improvements and potential regressions, reinforcing that multi-round repair pipelines require careful evaluation to avoid trading one weakness for another. Complementary evidence from manual reviews also suggests that repeated review cycles can increase secure output ratios when the reviewer is security-aware~\cite{SchaadGB25}. These results motivate \ours{}'s design choice to embed mitigation guidance directly into the reasoning process rather than relying on unconstrained self-correction.

Several approaches injected external security knowledge into generation. PromSec~\cite{NazzalKKP24} optimizes prompts for secure code generation and reports an order-of-magnitude reduction in operational time while improving security outcomes. SecureCoder~\cite{ZhaoSHLGZL25} maps tasks to CWE-aligned security knowledge and reports more than 65\% of vulnerability mitigation by aligning generation with CWE guidance. RESCUE~\cite{ShiZ25} addresses noise in raw security documents by constructing a distilled knowledge base and hierarchical retrieval, reporting a 4.8-point average improvement. Related recipes ``secure-by-instruction'' proposed a systematic use of CWE documentation during training or data synthesis, but typically operate at the dataset or fine-tuning time rather than as lightweight inference-time controls~\cite{LiRYY25}. In contrast, \ours{} targets a portable inference-time mechanism: concise, task-specific CWE mitigation snippets, and language-aware safeguards.

Another direction aligns the model reasoning with the vulnerability semantics. Nong \etal~\cite{NongACHCC24} proposed vulnerability-semantics-guided prompting and reported large gains in vulnerability identification accuracy by steering multi-step reasoning toward root causes rather than surface patterns. Separately, self-planning decomposes coding intent into subproblems prior to implementation, improving the correctness of complex generation tasks~\cite{JiangDWFSLJJ24}. These methods motivate \ours{}'s reasoning-centric design; however, \ours{} differs from them by grounding the reasoning steps in explicit CWE mitigation actions and language-specific secure defaults.

Security behavior varies across languages and benchmarks, making cross-language evaluations essential for general claims. Recent multi-language analyses have reported uneven security and quality in LLM-generated code~\cite{KharmaCAD25,ShahidAR25}. Benchmarking has also shifted to joint functionality and security metrics. CWEval evaluates whether the generated programs are correct and secure, highlighting that passing tests does not imply security~\cite{PengCHYR25}. BaxBench evaluates backend generation under exploit-driven assessment and shows that security failures remain common under realistic threat models, even when functional correctness is achieved ~\cite{VeroMCRBJHV25}. SecureAgentBench extends the evaluation to realistic vulnerability scenarios and reports that explicit security instructions alone do not reliably yield secure codes ~\cite{ChenHLASYZTLLZHL25}. \ours{} is positioned within this landscape as a prompt-level control designed to generalize across models, languages, and evaluation settings by integrating mitigation knowledge into the generation rationale.

\section{Methodology}\label{sec:method}

\begin{figure}[!t]
\centering
\includegraphics[width=\linewidth]{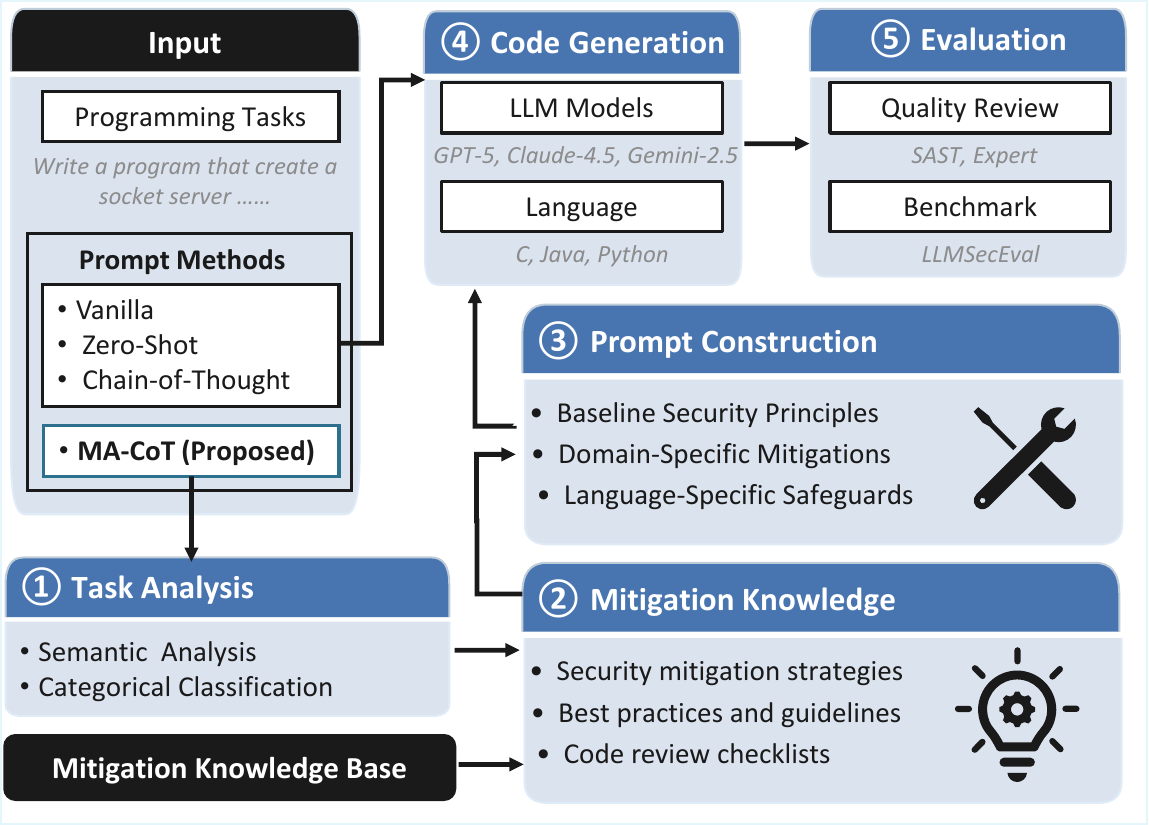}
\caption{Mitigation-Aware Chain-of-Thought (\ours{}) prompting method pipeline.}
\label{fig:pipeline}
\end{figure}

The proposed Mitigation-Aware Chain-of-Thought (MA-CoT) framework is designed to enhance the \textit{reliability} of LLM-generated code. As illustrated in Figure~\ref{fig:pipeline}, the framework operates via a multi-stage pipeline that progressively enriches prompts with security context and mitigation guidance. Each stage is intended to reduce recurring vulnerabilities and ensure consistent, dependable code generation across models, languages, and prompting strategies, thereby systematically improving the reliability of LLM-based software synthesis.

\subsection{Pipeline Design}
\BfPara{Stage 1: Task Analysis and Domain Classification} For a given programming task description, \ours{} determines relevant security domains by referencing the predefined categories below. It uses LLM-based semantic analysis to classify the task and assign one or more categories as follows: \ding{182} {\em Secure Coding.} Tasks grounded in known software weaknesses based on MITRE's CWE taxonomy, e.g., injection flaws, cryptographic misuse, and insecure data handling. \ding{183} {\em Data Structures and Algorithms.} Problems focusing on core computing concepts such as array manipulation, recursion, sorting, searching, and graph traversal. \ding{184} {\em Parsing and Validation.} Tasks requiring careful input processing, constraint enforcement, and boundary checking to assess robustness and correctness. \ding{185} {\em Networking.} Scenarios emulating networked systems, e.g., authentication errors and protocol misuse. \ding{186} {\em Mathematics and Logic.} Challenges evaluating numerical reasoning, precision, and logical flow. \ding{187}{\em Programming Systems and Utilities.} Utility tasks involving file handling, scripting, system configuration, and general-purpose workflows. \ding{188} {\em Concurrency and Synchronization.} Tasks targeting race conditions, thread safety, and synchronization in parallel execution tasks.

\BfPara{Stage 2: Mitigation Retrieval} For each identified domain, the pipeline queries the code-security weakness mitigation dataset in Section~\ref{sec:MA_DS} to retrieve mitigation strategies for the task. This database contains mitigation entries mapped to security categories and annotated with general rules, language-agnostic guidance, code review checklists, and/or fine-tuning examples. For example, authentication-related tasks retrieve mitigation for Weak Password Requirements: ``Implement strong password policies, including requirements for length, complexity, and expiration. Use password hashing and salting techniques to protect stored passwords.''

\BfPara{Stage 3: Prompt Construction} The final prompt combines three components: \ding{182} the baseline security principles applicable to all code, \ding{183} the mitigation strategies, and \ding{184} the specific security checklist for the task and the programming language based on the aforementioned Stage~2. This structured approach provides the LLM with knowledge related to secure code to improve the security of LLM-generated code.

\subsection{Task Dataset}
This work builds on the dataset in~\autoref{sec:data}, created to evaluate the security and quality of LLM-generated code. In addition, we developed a code weakness mitigation dataset capturing practical knowledge on how vulnerabilities should be mitigated rather than merely identified, translating it into a structure from which language models can directly learn.

\subsection{Experimental Design}\label{sec:expdesign}

\subsubsection{LLM, Language, and Prompt Selection}
To examine how securely LLMs generate code under different conditions, we narrowed our evaluation to a representative set of models, programming languages, and prompt styles reflecting common practice rather than idealized setups. The choices were guided by practicality and coverage, including tools developers rely on while capturing meaningful architectural variation.

This study focuses on three LLM families: ChatGPT, Gemini, and Claude. These models are widely deployed and rank among the strongest in code generation, user base, credibility, and support. They differ in training scale, context-window limits, and safety alignment strategies, often leading to behavioral differences when handling code mixing functionality with security requirements. \autoref{tab:model_configurations} lists the selected configurations. {\tt Max\_tokens} specifies the maximum tokens produced in completion. The temperature regulates output randomness. The CW is the largest token limit processed in a single forward pass, including prompt and generated responses, dictating how much text the LLM can handle at once.

\begin{table}[t]
\centering
\caption{Configurations: model, short name (SN), temperature (Temp), max tokens (MT), and context window (CW).}
\label{tab:model_configurations}\vspace{-1mm}
\scalebox{0.81}{
\begin{tabular}{lcccc}
\hline
Model & SN &  Temp & MT  & CW \\
\hline
gpt-5-2025-08-07~\cite{GPT5OpenAI25} & \gpt & 1 & 32k  &  400k  \\
claude-sonnet-4-5-20250929~\cite{Claude25} & \claude  & 1 & 32k  & 200k  \\
gemini-2.5-pro~\cite{Gemini25}   & \gemini & 1 & 32k  &  1000k \\
\hline
\end{tabular}
}
\end{table}

\BfPara{Programming Languages}
The evaluation spanned three languages: C, Java, and Python, each with distinct risks and features, allowing us to observe how LLMs adapt across languages. C is a low-level, memory-unsafe language exposing buffer overflows, use-after-free, and pointer-related vulnerabilities. Manual memory management requires careful discipline. Java is a managed language with garbage collection and strong typing, shifting vulnerabilities toward logic errors, injection flaws, and improper API usage. Python is a dynamically typed, interpreted language with an extensive standard library, where vulnerabilities involve input validation, insecure deserialization, and cryptographic misuse. Evaluating all three provides a balanced view of how models handle low- and high-level security concerns and reflects real-world software diversity.

\BfPara{Prompting Methods}
We compare four prompting methods reflecting common LLM use: \ding{182} Vanilla prompting provides only the task description without security guidance as a baseline. \ding{183} Zero-shot prompting introduces minimal structure by encouraging secure reasoning as ``Write secure code for the following task''. \ding{184} Chain-of-Thought (CoT) extends this by requesting step-wise reasoning: (1) Understand requirements, (2) Identify security considerations, (3) Implement the solution securely, and (4) Verify correctness. Although this improves correctness, it may overlook minor security concerns. \ding{185} The mitigation-aware CoT prompt, proposed in this work, integrates security signals into the reasoning stage through baseline rules, task-specific mitigation, and a language checklist. Rather than correcting code after generation, it encourages models to surface potential weaknesses during reasoning.
\subsubsection{Code Generation and Collection}
For each combination of (LLM, language, prompting method, task), we generated one code sample, resulting in 7{,}200 total implementations (3 LLMs $\times$ 3 languages $\times$ 4 methods $\times$ 200 tasks). The generation used API access to the respective model endpoints with consistent parameters. The output code was extracted, compiled/syntax-checked, and stored with associated metadata (model, language, prompt method, task ID).

\subsection{Evaluation Methodology}

We used SonarQube as our main automated analysis platform~\cite{SonarQube24}. It provides language-specific analyzers, and for each generated code sample we collected detected security vulnerabilities and their severity (Blocker, Critical, Major, Minor). We also mapped vulnerabilities to corresponding CWEs to identify common weakness types.

\subsection{Benchmarking}
To examine how the proposed prompting method influences code security, we used LLMSecEval (Section~\ref{LLMSecEval_DS}) as an external reference. This benchmark centers on coding scenarios where security failures are likely, rather than tasks focused only on functional output. This distinction is critical: models may produce code that compiles and behaves as expected yet includes unsafe operations, weak validation, or designs contradicting secure-coding practices. LLMSecEval captures these issues through tasks mapped to common CWE patterns, offering a clearer view of how models handle vulnerability-prone situations. The benchmark also fits the multi-language setup of our study. Because our experiments cover C, Java, and Python, a shared set of security-oriented tasks helps maintain consistency across languages with different programming paradigms and risk profiles. LLMSecEval provides consistency by framing each task to highlight the underlying security concerns rather than the surface-level syntax.

\subsection{Language-Aware Vulnerability Attribution}
\label{sec:pl_aware_attribution}

A recurring source of confusion in language-aware security analysis is the term \emph{language feature}. Many vulnerability drivers are not properties of the programming language specification itself, but of the \emph{language stack} that developers typically use: standard libraries, primary third-party libraries, platform APIs, and compiler toolchains. To avoid over-claiming, we use a strict, layered attribution scheme that separates \emph{language-core} mechanisms from \emph{language-adjacent} mechanisms.

\BfPara{Attribution Layers}
We label an attributed mechanism as \emph{language-core} only when it follows directly from the language specification or mandatory runtime semantics (\eg C raw buffers without bounds checks, Python default-argument evaluation timing). All other drivers are classified as language-adjacent and attributed to the closest layer: (1) standard runtime/library defaults, (2) ecosystem library/framework API design, (3) platform/OS API design, (4) toolchain/optimization behavior, or (5) application-level security logic/policy. This distinction is essential; e.g., TOCTOU (CWE-367) is an OS/POSIX check-use issue that can occur in any language using the same syscalls, not a C language feature in the strict sense.

\BfPara{Attribution Procedure}
For each SonarQube-reported weakness, we use the standardized analysis fields (language\_feature, feature\_category, feature\_mechanism, and comparative\_analysis) to assign one primary layer. When the same vulnerability could plausibly be attributed to multiple layers, we apply a conservative tie-break rule: we assign it to the \emph{lowest} layer that is necessary to produce the risk (\eg if the risk exists because a library API allows insecure options, we assign it to the library layer rather than to the language). This results in a reproducible mapping from observed vulnerabilities to language-core versus language-adjacent mechanisms, enabling a clearer interpretation of RQ2. Table~\ref{tab:layered_attribution} summarizes the attribution layers and the strict criteria used to avoid mislabeling OS, library, or toolchain mechanisms as language-core features.

\begin{table}[t]
\centering
\caption{Layered attribution used to interpret "language-aware" vulnerability drivers.}
\label{tab:layered_attribution}
\small
\begin{tabular}{p{0.97\columnwidth}}
\hline
\textbf{Language-core (strict):} Properties required by the language spec or mandatory runtime semantics (\eg C raw memory and unchecked buffers; Python default-argument evaluation; Python string literal semantics relevant to header injection). \\
\textbf{Standard runtime/library:} Defaults and APIs shipped with the standard platform (\eg Java JCE/JSSE/JAXP configuration surfaces). \\
\textbf{Ecosystem library/framework:} Third-party or dominant frameworks and their secure-by-default posture (\eg crypto/TLS libraries, web frameworks). \\
\textbf{Platform/OS API:} System call semantics and filesystem/network primitives (\eg POSIX check-then-use patterns driving TOCTOU). \\
\textbf{Toolchain/optimization:} Compiler/linker/runtime optimizations that affect security idioms (\eg non-guaranteed memory wiping without a secure zeroization primitive). \\
\textbf{Application security logic/policy:} Input validation, password policies, and configuration choices that are not determined by the language itself. \\
\hline
\end{tabular}
\end{table}

\section{Dataset}\label{sec:data}

\subsection{Prompt Description}
This study builds on our dataset~\cite{KharmaCAD25} developed to evaluate the functional quality and security of LLM-generated code. The dataset contains 200 language-agnostic programming tasks organized into seven functional categories. Each task includes a neutral problem statement, and generated code was evaluated using unit tests and manual and automated (tool-based) reviews, such as SonarQube. This dataset also serves as the basis for the prompt descriptions used in this study.

The reuse of this dataset offers two advantages. First, it preserves continuity with our earlier analysis, enabling us to assess whether changes in prompting or model capabilities yield measurable improvements in secure code generation. Second, because the dataset was reviewed and refined by multiple practitioners, it remains suitable for evaluating the prompt methods explored here, including Vanilla, Zero-shot, CoT, and the mitigation-aware CoT approach. We selected three LLMs and instructed them to generate code from this dataset using four prompting methods across three programming languages, producing in 7{,}200 samples for evaluation.

\subsection{Mitigation-Aware Dataset}
\label{sec:MA_DS}
While the first dataset evaluated model outputs, the second serves a complementary purpose. It captures practical knowledge on how vulnerabilities should be mitigated rather than merely identified, translating it into a structure from which language models can learn. Each record represents a security category or specific CWE identifier, linking a problem domain to language-agnostic or language-specific mitigation guidance, review checklists, and, in some cases, fine-tuning examples.

\BfPara{Dataset Composition}
Each entry in the Mitigation-Aware dataset contains four core components.  
First, \textit{General Rules} outline baseline security practices across languages, emphasizing input validation, secret handling, cryptography, TLS configuration, and concurrency safeguards.  
Second, \textit{Language-Specific Guidelines} tailor this guidance to languages such as C, C++, Python, and Java, highlighting defensive techniques like safe memory handling in C or secure resource management in Java.  
Third, the \textit{Code Review Checklist} provides a concise reference to confirm that generated code meets essential safety requirements.  
Finally, the dataset includes a few \textit{Fine-tuning Examples}, each in a three-part format: an instruction, an insecure input, and a secure output, illustrating correct remediation steps for models to learn. We include representative entries in the appendix. These examples show how each record integrates general mitigation principles, language-specific guidance, review checklists, and fine-tuning examples. They demonstrate the hierarchical organization of the Mitigation-Aware dataset and how security concepts are translated into practical, language-aware rules suitable for evaluation and model alignment.

\BfPara{Purpose and Applications}
The Mitigation-Aware dataset operates at two complementary levels. First, it serves as a knowledge base combining mitigation strategies, language-aware guidance, and reasoning that supports secure-programming decisions. This structure helps trace how coding practices reduce risk, especially when a vulnerability appears correct in isolation but fails upon review. Second, it acts as a broader training corpus. Although it lacks explicit input-output pairs, its rules and rationales shape model behavior by encouraging secure design patterns. This dual role supports both model analysis and security-aware tuning. The modular structure further allows additional languages, evolving weaknesses, or CWE domains to be incorporated with minimal disruption as security expectations change.

\subsection{LLMSecEval Benchmark Dataset}\label{LLMSecEval_DS}

LLMSecEval~\cite{tony2023llmseceval} is widely used to evaluate language models in security-sensitive coding tasks~\cite{TonyFMDS25, BruniGGK25, SiddiqSDM24, LiuSJKGMAN24}. The benchmark is built from prompts derived from real C and Python examples associated with the MITRE CWE Top~25, rewritten in neutral language so the task intent remains clear without revealing the vulnerability. Each prompt is paired with a secure reference solution, enabling consistent comparison of model outputs and highlighting meaningful gaps in secure reasoning. The dataset is particularly useful for evaluating prompt methods, as many of its tasks resemble routine programming problems in which security issues often go unnoticed. By covering a range of weakness categories rather than a single CWE, LLMSecEval helps to expose recurring model errors and provides a stable baseline for measuring improvement. This makes it a practical benchmark for assessing whether a prompt strategy genuinely enhances a model's ability to avoid risky patterns and follow safer design choices.

\section{Results and Discussion}\label{sec:results}

We present findings from evaluating \ours{} on three LLMs (\gpt, \claude, \gemini), three programming languages (C, Java, Python), and two benchmark datasets, totaling 350 security-relevant programming tasks. The analysis combines automated static analysis using SonarQube with manual expert review to assess vulnerability patterns, severity distributions, and the effectiveness of mitigation-aware prompting compared to other approaches.

\subsection{Performance on Primary Dataset}

Table~\ref{tab:our_dataset_results} presents the vulnerability counts from the 200-task dataset across all evaluated configurations. The results indicate substantial variation in baseline security behavior and the way different models respond to prompting interventions.

\begin{table}[t]
\centering
\caption{Number of vulnerabilities by programming language, model, and prompting strategy (200-task dataset).}
\label{tab:our_dataset_results}\vspace{-1mm}
\scalebox{0.87}{
\begin{tabular}{llcccc}
\hline
Language & Model & Vanilla & ZeroShot & CoT & \ours{} \\
\hline
\multirow{3}{*}{C} 
& \gpt      & 17 & 23 & 24 & 13 \\
& \claude & 19 & 20 & 27 & 11 \\
& \gemini & 16 & 20 & 16 & 7 \\
\hline
\multirow{3}{*}{Java} 
& \gpt      & 2 & 3 & 3 & 2 \\
& \claude & 20 & 13 & 12 & 1 \\
& \gemini & 5 & 6 & 6 & 1 \\
\hline
\multirow{3}{*}{Python} 
& \gpt      & 3 & 2 & 2 & 2 \\
& \claude & 7 & 10 & 4 & 2 \\
& \gemini & 3 & 3 & 4 & 0 \\
\hline
\end{tabular}}
\end{table}

Several patterns emerged from this evaluation. First, \ours{} consistently produced the fewest vulnerabilities in seven of the nine configurations. The exceptions occur with \gpt in Java, where Vanilla and \ours{} both generate two vulnerabilities, and with \gpt in Python, where ZeroShot, CoT, and \ours{} all generate two vulnerabilities. Second, the ZeroShot and CoT prompting methods frequently fail to improve security over the Vanilla baseline, except in Java and partially in Python for \claude. In C, ZeroShot and CoT increase the vulnerability counts for all three models, with CoT generating 27 issues for \claude compared to 19 in Vanilla. This surprising result suggests that encouraging step-by-step reasoning without concrete security knowledge can degrade code security.

\begin{takeaway}
\ours{} delivers the most consistent security gains, whereas ZeroShot and CoT often fail to help and can even increase vulnerabilities, particularly in C, suggesting that unguided reasoning may unintentionally weaken code security.\end{takeaway}

The magnitude of improvement in \ours{} varied by language. C shows clear gains, reducing vulnerabilities from 52 (vanilla) to 31 (\ours{}), a 40\% decrease. Java shows more dramatic results, dropping from 27 to 4 vulnerabilities (85\% reduction). Python achieved similar success, with a 69\% reduction from 13 to 4 issues. These differences reflect how cryptographic algorithms, memory management, and certificate validation affect baseline code security.

\claude shows notable behavior. It produces the highest Vanilla vulnerability count (46 across all languages), but achieves the largest relative improvement under \ours{}, reducing issues to 14 in total. This 70\% reduction indicates a strong instruction-following capability paired with weak security priors. The model appears to lack internal safeguards that would prevent vulnerabilities in unconstrained generation but can effectively apply explicit mitigation guidance when provided with such safeguards.

\begin{takeaway}
\claude demonstrates limited built-in security safeguards, yet it shows strong responsiveness to structured mitigation prompts, resulting in a marked reduction in vulnerability cases under guided conditions.
\end{takeaway}

\subsection{Benchmarking using LLMSecEval}

Table~\ref{tab:llmseceval_results} presents the results of the LLMSecEval benchmark, which contains 150 security-focused tasks designed independently from our primary dataset. This external validation tests whether \ours{} generalizes beyond the specific characteristics of the evaluation dataset.

\begin{table}[t]
\centering
\caption{Vulnerabilities by programming language, model, and prompting strategy (LLMSecEval, 150 tasks).}
\label{tab:llmseceval_results}\vspace{-1mm}
\scalebox{0.87}{
\begin{tabular}{llcccc}
\hline
Language & Model & Vanilla & ZeroShot & CoT & \ours{} \\
\hline
\multirow{3}{*}{C} 
& \gpt      & 4 & 7 & 10 & 0 \\
& \claude & 3 & 3 & 7 & 0 \\
& \gemini & 4 & 5 & 1 & 1 \\
\hline
\multirow{3}{*}{Java} 
& \gpt      & 2 & 0 & 0 & 0 \\
& \claude & 46 & 28 & 6 & 2 \\
& \gemini & 5 & 3 & 3 & 1 \\
\hline
\multirow{3}{*}{Python} 
& \gpt      & 1 & 0 & 1 & 0 \\
& \claude & 4 & 2 & 3 & 0 \\
& \gemini & 4 & 6 & 3 & 0 \\
\hline
\end{tabular}}
\end{table}

The LLMSecEval results validated the patterns observed in our primary dataset. \ours{} achieved complete vulnerability elimination in six of the nine configurations. When vulnerabilities persist, they remain minimal (one or two instances) and outperform all other prompting approaches. The method is effective even for tasks explicitly constructed to expose weaknesses in automated code generation.

A comparison of the two datasets shows both consistency and instructive differences. The relative ranking of the prompting methods remained stable: \ours{} consistently outperformed the alternatives, while ZeroShot, CoT, and Vanilla showed no consistency in their security quality. However, the absolute vulnerability counts differed substantially. LLMSecEval produces a higher vulnerability ratio using Vanilla in Java than our dataset (53 versus 27 in our dataset), but lower C counts (11 versus 52). These variations reflect differences in tasks and programming languages rather than evaluation methods. LLMSecEval emphasizes web security patterns common in Java applications, while our dataset includes more system programming scenarios typical of C development.

\begin{takeaway}
\ours{} is consistently the most secure prompting method: it removes vulnerabilities in most settings, and when issues remain, they are few and fewer than all alternatives. Across both datasets, \ours{} maintained the top rank, while ZeroShot, CoT, and Vanilla were unstable. Differences in raw vulnerability counts stem mainly from task types and language focus, not evaluation flaws.\end{takeaway}

The consistent and counterintuitive CoT behavior across both datasets warrants closer examination. In C, CoT yields higher vulnerability rates than Vanilla in five model-dataset pairs. When given general reasoning instructions without explicit security guidance, models fail to avoid vulnerabilities that are sometimes prevented in Vanilla and ZeroShot settings.

\subsection{Vulnerability Severity Analysis}
\label{sec:vuln_severity}

\begin{figure*}[t]
\centering
\includegraphics[width=\textwidth]{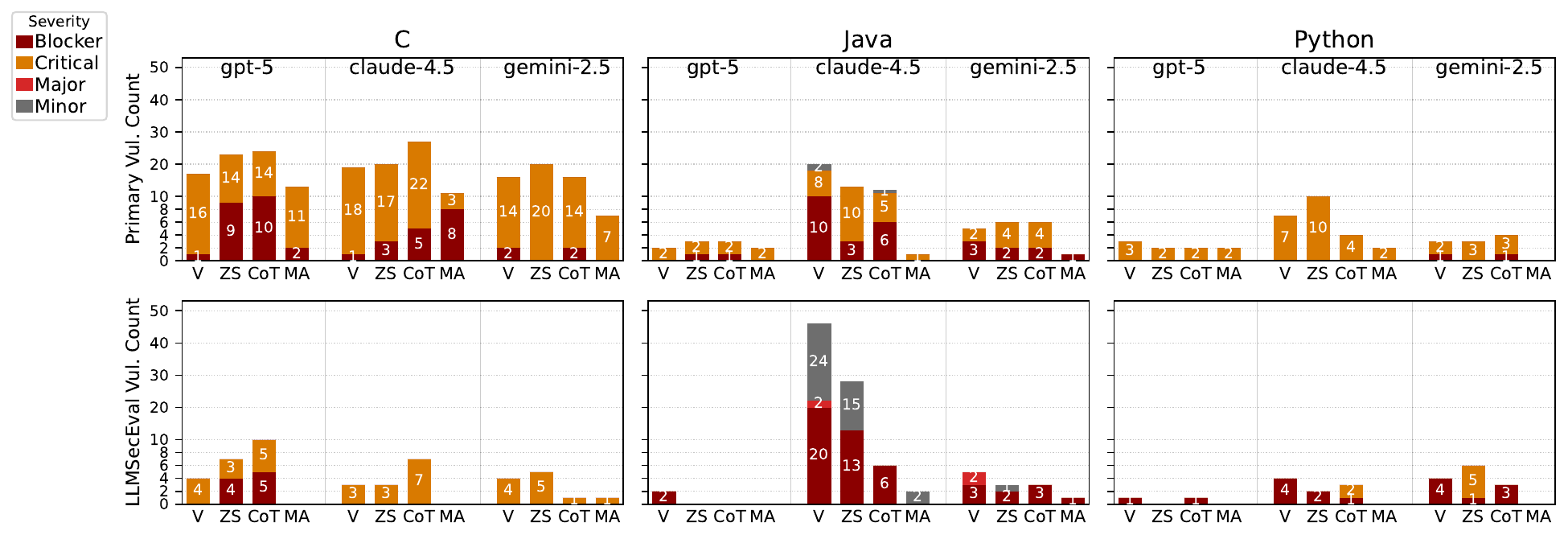}
\caption{Vulnerability severity counts across two datasets. Panels are organized by dataset (rows) and language (columns). Within each panel, bars are grouped by LLM and ordered by prompting method (V=Vanilla, ZS=Zero-shot, CoT=Chain-of-Thought, MA=\ours{}). Stacks show severity counts (Blocker, Critical, Major, Minor).}
\label{fig:severity_two_datasets}
\end{figure*}

Severity reflects the likely impact of the identified vulnerabilities and provides more insight than raw counts alone. SonarQube categorizes each issue into one of four levels (Blocker, Critical, Major, Minor). We present severity distributions by language, LLM, and prompting strategy in~\autoref{fig:severity_two_datasets}.

\BfPara{Primary Dataset}
The findings were concentrated almost exclusively at the two highest severity levels. In the Vanilla setting, we observed 92 total findings, of which 90 were Blocker or Critical (18 Blocker, 72 Critical, 2 Minor). With \ours{}, the total decreases to 39 findings, all of them high-severity (11 Blocker, 28 Critical), corresponding to a 56.7\% reduction in Blocker and Critical issues (from 90 to 39). This improvement is consistent across languages: in Java, high-severity findings fall from 25 (13 Blocker, 12 Critical) to 4 (1 Blocker, 3 Critical), and in Python from 13 (1 Blocker, 12 Critical) to 4 (0 Blocker, 4 Critical). C shows a category shift under \ours{}: Critical findings drop from 48 to 21, while Blocker increases from 4 to 10, the combined high-severity total (Blocker plus Critical) still decreases from 52 to 31. The Blocker increase is dominated by \SonarRule{c:S5798} (\emph{``\texttt{memset} should not be used to delete sensitive data''}). This rule flags a specific insecure-erasure idiom, using \texttt{memset()} to scrub a buffer immediately before releasing it, because the compiler may optimize the write operation. In other words, \ours{} reduces the overall high-severity burden in C, but it also introduces more \emph{attempted} buffer-sanitization steps that must be implemented with a guaranteed-wipe primitive to be correct. Therefore, we treat \SonarRule{c:S5798} as an implementation-level hardening issue rather than evidence of weaker security, and document the remediation in the appendix.

\BfPara{Benchmark Dataset (LLMSecEval)}
The benchmark shows a broader severity distribution than the primary dataset, including non-zero Major and Minor findings. Under Vanilla, we recorded 73 findings in total (34 Blocker, 11 Critical, 4 Major, 24 Minor), of which 45 were high-severity (Blocker+Critical). \ours{} reduces this to 4 findings (1 Blocker, 1 Critical, 0 Major, 2 Minor), reducing high-severity by 95.6\% (45 to 2) and total findings by 94.5\% (73 to 4). The same pattern holds across languages: in C, Critical decreases from 11 to 1; in Java, Blocker decreases from 25 to 1, and Major is eliminated (4 to 0); and in Python, Blocker decreases from 9 to 0, with no remaining findings under \ours{}.

\BfPara{Cross-Dataset Interpretation}
\autoref{fig:severity_two_datasets} provides four comparisons: (1) severity composition for a fixed configuration (stack segments), (2) differences between prompting methods within each LLM (bar order per group), (3) variation across languages (C, Java, Python), and (4) stability across datasets (Primary on top, LLMSecEval on bottom). In the primary dataset, findings concentrate in Blocker and Critical, whereas LLMSecEval includes Major and Minor findings. \ours{} reduces Blocker and Critical in both datasets and also lowers low-severity findings in LLMSecEval. We report per-category counts rather than totals to avoid masking shifts between severity levels and to keep improvements visible per language, model, and prompting method.
\begin{takeaway}
\ours{} lowers high-severity vulnerabilities (Blocker and Critical) across both datasets: by 56.7\% in the primary dataset (90 to 39) and by 95.6\% in LLMSecEval (45 to 2). On LLMSecEval, it also reduces total findings by 94.5\% (73 to 4), indicating robustness under dataset shift.\end{takeaway}
\begin{takeaway}
In C of the primary dataset, \ours{} reduces Blocker and Critical (52 to 31) while increasing Blocker due to strict secure-erasure enforcement, dominated by \SonarRule{c:S5798}. This reflects increased use of hardening patterns implemented with a non-guaranteed primitive (plain \texttt{memset}), not weaker security: the net high-severity burden decreases, and remaining issues are concentrated in a small set of cases.\end{takeaway}

\subsection{Common Weakness Enumeration Patterns}
\label{subsec:cwe_patterns}
\begin{figure*}[t]
\centering
\includegraphics[width=\textwidth]{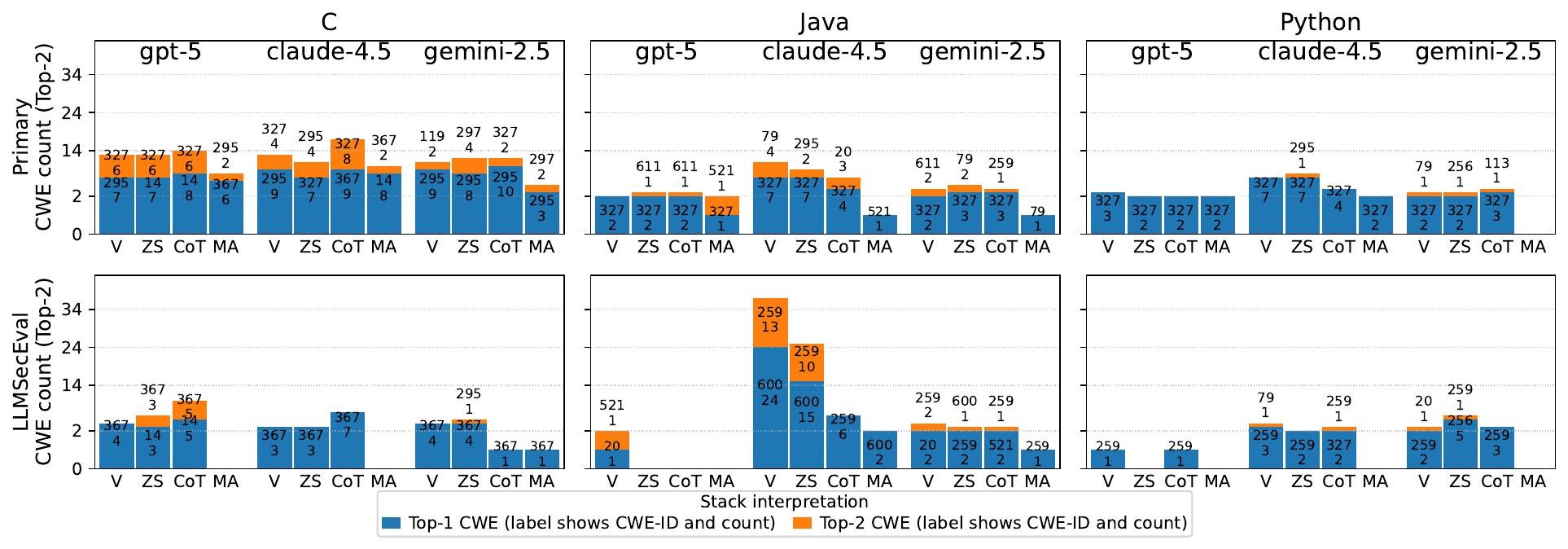}
\caption{Common Weakness Enumeration (CWE) patterns across datasets, languages, LLMs, and prompting methods. Rows correspond to datasets (top: Primary, bottom: LLMSecEval). The columns correspond to the languages (C, Java, and Python). Within each panel, bars are grouped by LLM and ordered by prompting method (V=Vanilla, ZS=Zero-shot, CoT=Chain-of-Thought, MA=\ours{}).}
\label{fig:cwe_patterns_top2}
\vspace{-1mm}
\end{figure*}

Severity counts quantify impact but do not explain recurring weakness classes. We therefore analyzed the Common Weakness Enumeration (CWE) categories reported by SonarQube to characterize recurring vulnerability types by dataset, language, LLM, and prompting method. \autoref{fig:cwe_patterns_top2} summarizes the two most frequent CWE categories per configuration, while \autoref{tab:cwe_patterns_primary_benchmark_top5} reports the top five CWEs, providing a fine-grained view of which classes \ours{} removes and which remain.

\begin{table}[t]
\centering
\caption{Top-5 CWE categories per language, LLM, and prompting method across datasets. Each cell lists up to five categories as CWE-ID (count), without the \texttt{CWE-} prefix.}
\label{tab:cwe_patterns_primary_benchmark_top5}\vspace{-1mm}
\scriptsize
\setlength{\tabcolsep}{2.2pt}
\renewcommand{\arraystretch}{1.05}
\scalebox{0.81}{
\begin{tabular}{p{0.09\columnwidth}ll|p{0.44\columnwidth}|p{0.35\columnwidth}}
\hline
Lang. & LLM & Method & Primary (Top-5 CWEs) & LLMSecEval (Top-5 CWEs) \\
\hline
\multirow{12}{*}{C} & \multirow{4}{*}{GPT-5} & V & 295(7), 327(6), 367(2), 119(1), 297(1) & 367(4) \\
 &  & ZS & 14(7), 327(6), 367(5), 119(2), 295(2) & 14(3), 367(3), 119(1) \\
 &  & CoT & 14(8), 327(6), 295(4), 367(3), 119(2) & 14(5), 367(5) \\
 &  & MA & 367(6), 295(2), 327(2), 14(1), 119(1) & -- \\
\cline{2-5}
 & \multirow{4}{*}{Claude-4.5} & V & 295(9), 327(4), 297(3), 367(2), 119(1) & 367(3) \\
 &  & ZS & 327(7), 295(4), 367(4), 119(3), 297(2) & 367(3) \\
 &  & CoT & 367(9), 327(8), 14(5), 295(4), 297(1) & 367(7) \\
 &  & MA & 14(8), 367(2), 295(1) & -- \\
\cline{2-5}
 & \multirow{4}{*}{Gemini-2.5} & V & 295(9), 119(2), 297(2), 327(2), 367(1) & 367(4) \\
 &  & ZS & 295(8), 297(4), 367(4), 120(2), 327(2) & 367(4), 295(1) \\
 &  & CoT & 295(10), 327(2), 119(1), 297(1), 367(1) & 367(1) \\
 &  & MA & 295(3), 297(2), 367(2) & 367(1) \\
\hline
\multirow{12}{*}{Java} & \multirow{4}{*}{GPT-5} & V & 327(2) & 20(1), 521(1) \\
 &  & ZS & 327(2), 611(1) & -- \\
 &  & CoT & 327(2), 611(1) & -- \\
 &  & MA & 327(1), 521(1) & -- \\
\cline{2-5}
 & \multirow{4}{*}{Claude-4.5} & V & 327(7), 79(4), 20(3), 600(2), 611(2) & 600(24), 259(13), 20(9) \\
 &  & ZS & 327(7), 79(2), 295(2), 521(2) & 600(15), 259(10), 79(2), 521(1) \\
 &  & CoT & 327(4), 20(3), 259(2), 79(1), 521(1) & 259(6) \\
 &  & MA & 521(1) & 600(2) \\
\cline{2-5}
 & \multirow{4}{*}{Gemini-2.5} & V & 327(2), 611(2), 521(1) & 20(2), 259(2), 521(1) \\
 &  & ZS & 327(3), 79(2), 521(1) & 259(2), 600(1) \\
 &  & CoT & 327(3), 79(1), 259(1), 323(1) & 521(2), 259(1) \\
 &  & MA & 79(1) & 259(1) \\
\hline
\multirow{11}{*}{Python} & \multirow{4}{*}{GPT-5} & V & 327(3) & 259(1) \\
 &  & ZS & 327(2) & -- \\
 &  & CoT & 327(2) & 259(1) \\
 &  & MA & 327(2) & -- \\
\cline{2-5}
 & \multirow{4}{*}{Claude-4.5} & V & 327(7) & 259(3), 79(1) \\
 &  & ZS & 327(7), 295(1), 297(1), 521(1) & 259(2) \\
 &  & CoT & 327(4) & 327(2), 259(1) \\
\cline{2-5}
 & \multirow{3}{*}{Gemini-2.5} & V & 327(2), 79(1) & 259(2), 20(1), 79(1) \\
 &  & ZS & 327(2), 256(1) & 256(5), 259(1) \\
 &  & CoT & 327(3), 113(1) & 259(3) \\
  &  & MA & -- & -- \\
\hline
\end{tabular}
}
\normalsize
\end{table}

\BfPara{Primary Dataset}
In C, baseline prompting is marked by certificate-validation and cryptography issues, mainly CWE-295 (Improper Certificate Validation) and CWE-327 (Broken or Risky Cryptographic Algorithm). Aggregated across LLMs, these remain frequent under Vanilla, Zero-shot, and CoT (CWE-295: 25, 14, 18; CWE-327: 12, 15, 16). Under \ours{}, both drop sharply (CWE-295=6; CWE-327=2). Residual findings concentrate on CWE-367 (TOCTOU Race Condition) and CWE-14 (Compiler Removal of Code to Clear Buffers), linked to specific idioms: TOCTOU arises from check-then-use flows, while CWE-14 is triggered when a wipe such as \texttt{memset} is not guaranteed under compiler optimization (appendix).

The appearance of CWE-367 and CWE-14 under \ours{} requires careful interpretation. In our outputs, these reflect security hardening that triggers stricter static analysis. For example, file-existence checks before opening can create a time-of-check time-of-use window (CWE-367). Similarly, explicit buffer scrubbing with \texttt{memset} may be flagged because the compiler can optimize the write, motivating guaranteed-wipe primitives (appendix B). We therefore treat these residual findings as implementation-level hardening issues rather than baseline cryptographic or certificate-validation failures.

Java shows a different profile: CWE-327 persists under baseline alongside web and API-misuse patterns such as CWE-611 (Improper Restriction of XML External Entity Reference) and CWE-79 (Cross-site Scripting). Under \ours{}, CWE-611 is absent in the primary dataset, and CWE-79 is strongly reduced, indicating that explicit configuration-level guidance is most effective when security depends on correct API defaults rather than syntactic correctness alone.

Python shows the most concentrated weaknesses in the primary dataset, with CWE-327 leading in multiple baseline configurations. This aligns with Python’s memory safety by design, where security failures cluster around cryptographic misuse and unsafe parameter choices. Under \ours{}, Python findings are minimized and eliminated in some settings.

Cryptographic weaknesses (CWE-327) appear in all three languages, suggesting that secure cryptography remains challenging, regardless of the language design. In the primary dataset, the reductions in CWE-327 under \ours{} were substantial when measured against the Vanilla baseline: in C, 12 to 2 (83\%); in Java, 11 to 1 (91\%); and in Python, 12 to 4 (67\%). These reductions are consistent with \ours{} prompts that constrain algorithm choices and parameters toward modern, safe defaults. The appendix provides a task-level qualitative example for \claude, showing this shift from ECB/CBC-style implementations under baseline prompts to AEAD-based designs under \ours{} for the same encryption task.

\begin{takeaway}
\ours{} suppresses recurring high-impact CWE classes across languages and LLMs, indicating a \textbf{consistent reduction} in systematic security failure modes rather than only a drop in total results.
\end{takeaway}

\BfPara{LLMSecEval Benchmark}
The benchmark preserves the ranking observed in the primary dataset while altering the prevalent CWE mix for some languages. In C, findings are mainly race and timing issues (CWE-367), and \ours{} removes them in most configurations, leaving one residual case. In Java, the benchmark exposes uncaught exceptions and authentication and credential-management weaknesses (\ie CWE-600 and CWE-259), which \ours{} substantially reduces compared to baseline prompts. In Python, findings also concentrate on few categories, and \ours{} eliminates them across configurations. Overall, \ours{} not only lowers total counts but filters high-risk, frequent CWE categories, leaving a smaller set tied to specific implementations and easier to mitigate with focused hardening.

\subsection{Model-Specific Security Characteristics}

\autoref{tab:model_security} compares the three models by aggregating vulnerabilities in C, Java, and Python, reported separately for the Primary (200 tasks) and LLMSecEval (150 tasks) datasets. Because the datasets differ in size and composition, we interpret cross-dataset differences qualitatively and focus on within-dataset changes across prompting methods. Across both datasets, \ours{} was the only prompting strategy that consistently reduced vulnerability counts for every model. In contrast, generic security phrasing (Zero-shot) and basic reasoning (CoT) are not dependable: on the primary dataset, they increase findings for all models relative to Vanilla, and on LLMSecEval, they remain less effective than \ours{}.

\begin{table}[t]
\centering
\caption{Total vulnerabilities by model aggregated across languages. Prompting methods: V=Vanilla, ZS=Zero-shot, CoT=Chain-of-Thought, MA=\ours{}.}
\label{tab:model_security}\vspace{-1mm}
\setlength{\tabcolsep}{5pt}
\begin{tabular}{lcccccccc}
\hline
& \multicolumn{4}{c}{Primary Dataset} & \multicolumn{4}{c}{LLMSecEval Dataset}\\
\cmidrule(lr){2-5}\cmidrule(lr){6-9}
Model & V & ZS & CoT & MA & V & ZS & CoT & MA \\
\hline
\gpt    & 22 & 28 & 29 & 17 & 7  & 7  & 11 & 0 \\
\claude & 46 & 43 & 43 & 14 & 53 & 33 & 16 & 2 \\
\gemini & 24 & 29 & 26 & 8  & 13 & 14 & 7  & 2 \\
\hline
\end{tabular}
\vspace{-1mm}
\end{table}

\BfPara{Claude} \claude has the weakest baseline in the primary dataset (Vanilla=46) but the strongest improvement under \ours{} (14), a 70\% reduction. The same pattern appears in LLMSecEval (53 to 2, 96\%). This indicates high responsiveness to mitigation-aware constraints: when guidance specifies safeguards, the model produces measurably safer code.

\BfPara{Gemini} \gemini achieves the lowest \ours{} total in the primary dataset (8), improving from Vanilla=24 (67\%). In LLMSecEval, \ours{} reduced the findings from 13 to 2 (85\%). In particular, \gemini also shows less degradation under CoT in the reference dataset (7) than in the primary dataset (26), suggesting that its performance is more sensitive to task composition than to model capacity alone.

\BfPara{ChatGPT} \gpt is comparatively stable but less responsive in the primary dataset: Vanilla, Zero-shot, and CoT are close (22, 28, 29), and \ours{} reduces to 17 (23\%). In LLMSecEval, however, \ours{} eliminates all detected findings. This suggests that mitigation-aware prompting effectiveness depends on the task vulnerability mix.

These results support two conclusions: First, model selection and prompt design should be considered jointly; a weaker baseline model can still be effective when mitigation-aware prompting is consistently applied. Second, relying on generic ``be secure'' or reasoning prompts alone is risky; structured, actionable mitigations are needed for reliable security improvements across models and datasets.

\begin{takeaway}
Prompt design is a first-order factor: across models and datasets, \ours{} is the only method that consistently reduces vulnerabilities, while Zero-shot and CoT are unreliable and can increase findings relative to Vanilla.
\end{takeaway}

\begin{takeaway}
The results reveal an interaction between \emph{model behavior} and \emph{task vulnerability mix}. \ours{} is most effective when dominant weaknesses arise from API configuration and mitigation choices (\eg cryptography, credential handling), whereas residual findings concentrate on implementation-level hardening (\eg race conditions, secure wiping). This explains why the same model shows modest gains on one dataset and near elimination on another under the same prompting strategy.\end{takeaway}

\subsection{Language-Core/Stack Vulnerabilities Drivers}
\label{sec:pl_drivers_strengthening}
This subsection addresses RQ2 using a strict definition of \emph{language-specificity}. We separate (i) \emph{language-core} drivers (required by the language specification or mandatory runtime semantics) from (ii) \emph{language-stack} drivers, including standard runtime, ecosystem libraries/frameworks, platform/OS APIs, and toolchains. The attribution layers are defined in Table~\ref{tab:layered_attribution}.

\BfPara{Cross-Language Distribution of Drivers}
Table~\ref{tab:layer_contrib} summarizes each layer’s contribution to attributed weakness instances per language, with primary CWE evidence. The key result is that C security outcomes are shaped mainly by the surrounding stack: OS-level check-then-use filesystem patterns (TOCTOU, CWE-367) and library-level crypto/TLS configuration freedom, with language-core memory/bounds issues forming a smaller but important component. Java and Python shift toward exception boundaries, secret representation, configuration surfaces, and permissive APIs. This explains why \ours{} is strongest in Java and Python, where weaknesses are largely configuration/API-driven, while C retains residual risk from OS/toolchain semantics and language-core memory issues.

\begin{table}[t]
\centering
\caption{Layer contributions per language with the dominant CWE signal per layer. Percentages are computed over the attributed weakness instances for each language.}
\label{tab:layer_contrib}
\small
\begin{tabular}{c|p{0.76\columnwidth}}
\hline
 Lang. & Layer Contribution \\
\hline
 C & 
 Platform/OS API (38.0\%, dominated by CWE-367); Ecosystem crypto/TLS library API (31.2\%, dominated by CWE-327 and certificate validation CWEs). Language-core memory and bounds (27.1\%, dominated by CWE-119/120). Toolchain/optimization (1.8\%, dominated by CWE-14). Remaining cases are sparse or uncategorized. \\
 \hline
Java & Application security logic/policy (27.6\%, input validation, output encoding, auth configuration). Language-core error model exposure (27.1\%, dominated by CWE-600). Standard runtime crypto/TLS surfaces (21.2\%, dominated by CWE-327). Language-core secret representation gaps (13.5\%, dominated by CWE-259). XML/runtime configuration surfaces and other categories contribute the remainder. \\
\hline
Python & Generation-prone patterns and community idioms (30.8\%, \eg placeholder constants and repetition). Language-core semantics (21.5\%, \eg default-argument timing and string literal behavior relevant to injection). Application security logic/policy (21.5\%). Ecosystem library/framework API permissiveness (16.9\%, dominated by CWE-327/329 patterns). Runtime/module-state exposure (9.2\%). \\
\hline
\end{tabular}
\end{table}

\BfPara{C: Separating language-core memory hazards from, library, OS, and toolchain exposures}
C shows a split between language-core and language-stack risk. The language-core driver is the lack of intrinsic bounds checking with pervasive raw buffers and pointer arithmetic, consistent with memory-safety CWEs (CWE-119/120) in~\autoref{tab:layer_contrib}. In contrast, several high-frequency issues are not strictly C-specific (\autoref{tab:layered_attribution}): (1) TOCTOU (CWE-367) stems from POSIX check-then-use semantics and can arise in any language using the same OS APIs. (2) Cryptographic and TLS weaknesses are driven by library configuration surfaces (\eg algorithm/mode selection, verification settings) rather than C syntax. (3) Secure wiping failures (\eg CWE-14 and related rules) stem from toolchain optimization when non-guaranteed zeroization is used.

\BfPara{C Strengthening Actions by Layer}
The mitigation target depends on the layer (\autoref{tab:layer_contrib}):
\begin{enumerate*}
\item[\ding{172}] {\em Language-core (memory and bounds):} Prefer bounded abstractions (length-carrying buffers and checked wrappers), ban unbounded copies, and validate sizes at every boundary. Where feasible, adopt safer subsets (\eg CERT C / MISRA guidance) and enable sanitizers in testing to expose overflow- and lifetime-related defects early.
\item[\ding{173}] {\em Platform/OS API (TOCTOU):} Avoid check-then-use flows and prefer atomic, descriptor-based patterns (open-and-check rather than pre-check, fd (file descriptor)-based validation). Avoid separate existence checks and use primitives that reduce race windows (\eg \texttt{openat}, \texttt{O\_CREAT| O\_EXCL}, and \texttt{O\_NOFOLLOW} where applicable).
\item[\ding{174}] {\em Crypto/TLS library layer:} Reduce insecure configuration freedom by using safe wrappers and secure defaults. In \ours{}, require certificate and hostname verification (no verification bypass) and constrain cryptography to modern AEAD with safe parameter handling.
\item[\ding{175}] {\em Toolchain/optimization layer (secure wiping):} Use guaranteed-wipe primitives rather than plain \texttt{memset}-based scrubbing that compilers may remove (\eg \texttt{memset\_s} where available, or platform-provided secure-zero functions such as \texttt{explicit\_bzero}/\texttt{SecureZeroMemory}).
\end{enumerate*}

\begin{takeaway}
Based on~\autoref{tab:layered_attribution}, many high-frequency C findings (\eg TOCTOU and crypto/TLS misconfiguration) are attributed to platform and library layers rather than the C language-core, while true language-core risk remains concentrated in memory and bounds handling (\autoref{tab:layer_contrib}).\end{takeaway}

\BfPara{Java: Security failures concentrate in exception boundaries and string-configured security APIs}
Java’s managed runtime reduces memory-corruption weaknesses, shifting risk toward configuration-heavy security and error-driven failures. In~\autoref{tab:layer_contrib}, Java shows a strong error-handling contribution dominated by CWE-600, indicating that LLM-generated code often mishandles exceptional control flow (\eg uncaught exceptions, overly broad catches, error propagation bypassing checks). A second signal is hardcoded secrets (CWE-259), as security-sensitive values are expressed as string literals without a first-class secret type. Cryptographic weaknesses (CWE-327) persist because Java exposes algorithm and mode choices through string-based configuration (\eg transformation strings, provider selection), making insecure options easy to select without guardrails.

\BfPara{Java Strengthening Actions by Layer}
The mitigation goal varies by layer:
\begin{enumerate*}
\item[\ding{172}] {\em Language-core error boundaries:} Enforce fail-closed exception handling, avoid catch-all swallowing, ensure security checks and authorization decisions are not bypassed on exceptional paths, and prevent sensitive internal error details from leaking through exception messages.
\item[\ding{173}] {\em Secret representation and lifetime:} Avoid embedding secrets as string literals or long-lived \texttt{String}s (which cannot be reliably cleared and are easily propagated). Prefer typed secret carriers where feasible (\eg key objects, keystores, or short-lived \texttt{byte[]} buffers with best-effort zeroization) and never log or include secrets in exception text.
\item[\ding{174}] {\em Standard runtime crypto/TLS surfaces:} Constrain algorithm selection to modern AEAD by default and require cryptographically secure randomness and standard KDFs for derived keys. For TLS, certificate and hostname verifications are required, and verification-bypass patterns are forbidden.
\item[\ding{175}] {\em Application/framework layer:} Use output encoding or safe templating defaults to prevent CWE-79, validate all untrusted inputs to reduce CWE-20, and enforce authentication and password policies (CWE-521) explicitly via configuration and centralized validation.
\end{enumerate*}

\BfPara{Python: Expressiveness amplifies configuration risk and parameter reuse}
Python’s memory safety reduces low-level memory-corruption risk, but cryptographic misuse and input-driven vulnerabilities (\eg insufficient validation, unsafe string construction) remain common. As shown in~\autoref{tab:layer_contrib}, Python weaknesses stem from permissive APIs and generation-prone idioms, alongside language-core semantics that make insecure patterns concise (\eg module-level state, default-argument evaluation reusing values across calls). Models often encode security-critical parameters as copyable constants (\eg static IVs/nonces, placeholder secrets), and many APIs accept them without enforcing freshness or entropy. This interaction motivates guidance requiring nonce/IV uniqueness, secure randomness, and non-literal secret sourcing.

\BfPara{Python Strengthening Actions by Layer}
The following actions target Python-specific risks across language and stack layers.
\begin{enumerate*}
\item[\ding{172}] {\em Language-core semantics and state:} Avoid module-level security-critical constants and avoid default-argument traps by requiring per-operation runtime initialization of nonces/IVs/tokens using a CSPRNG. Treat all external inputs as untrusted and apply validation/encoding at security-sensitive sinks (\eg HTML output, command execution, and query construction).
\item[\ding{173}] {\em Library/framework APIs:} Prefer high-level cryptographic APIs that enforce randomness and safe modes by construction; explicitly forbid insecure modes and static parameters (\eg fixed IVs/nonces). For web frameworks, require safe templating/auto-escaping by default and explicit content-type handling.

\end{enumerate*}

\begin{takeaway}
Java and Python weaknesses are characterized by configuration, exception boundaries, and API choices, so \ours{} is most effective when it constrains these surfaces with secure defaults and fail-closed policies. C retains more residual risk because a larger fraction is driven by OS/toolchain semantics and language-core memory hazards, which require atomic OS patterns and secure-by-construction primitives in addition to prompt guidance.
\end{takeaway}

\begin{takeaway}
Language-core rules are necessary but insufficient; secure generation must also include stack-level constraints (library defaults, OS-safe patterns, etc.).\end{takeaway}

\section{Discussion}\label{sec:discussion}
Our results suggest that LLM-generated code security depends less on "more reasoning" and more on whether prompts provide \emph{actionable} mitigations aligned with dominant weakness types. Across both datasets, \ours{} was the most secure method, while Zero-shot and CoT were unstable. This stability appears at the severity level: high-severity findings (Blocker plus Critical) drop from 90 to 39 in the primary dataset (56.7\%) and from 45 to 2 in LLMSecEval (95.6\%).

A key technical signal is that \ours{} reduces recurring, high-impact CWE classes rather than totals alone. In the primary dataset, C was dominated by CWE-295 and CWE-327, and \ours{} reduced both (CWE-295=6; CWE-327=2). Remaining issues shift toward implementation-level hardening (\ie TOCTOU and secure wiping). This is most apparent in C, where the Blocker increase is driven by \SonarRule{c:S5798}, flagging insecure-erasure with \texttt{memset()} before \texttt{free()} due to compiler optimization; the net high-severity burden still decreases (52 to 31). The appendix provides a concrete cross-language illustration of this mechanism for a representative encryption task.

\BfPara{Why generic CoT can increase findings (especially in C)}
The underperformance of generic CoT is not accidental. CoT expands intermediate design choices, and without concrete mitigation constraints, models often select insecure defaults while producing plausible explanations. In C, this effect is amplified because added planning steps introduce check-then-use file flows and low-level plumbing, where TOCTOU and verification bypass patterns appear. In contrast, \ours{} narrows these degrees of freedom by prescribing safe primitives and explicit prohibitions, reducing risky configurations.

\BfPara{Layered attribution explains the cross-language gap}
The layered attribution in~\autoref{sec:pl_aware_attribution} clarifies why improvements differ by language. In C, residual risk is largely driven by language-stack layers, including OS-level filesystem semantics (\eg TOCTOU) and toolchain behavior (\eg non-guaranteed wiping), which prompts alone cannot eliminate. In Java and Python, weaknesses are more configuration and API-driven (crypto choices, exception boundaries, secret handling), making them more responsive to \ours{} constraints that enforce secure defaults and fail-closed behavior.

Dataset shift explains why the same prompting strategy appears moderate in one dataset and nearly eliminated in another. LLMSecEval emphasizes different task families (\ie Java web patterns), whereas our dataset includes more systems-oriented C scenarios; differences in raw counts reflect task composition and language mix, not evaluation artifacts. Prompt effectiveness should be assessed by \emph{which CWE classes dominate} in a benchmark, not only aggregate totals.

Our study provides three practical and testable insights.
\ding{182} treat prompt design as a security control (\ours{} is consistently strongest, whereas generic CoT/Zero-shot is not reliable); \ding{183} report severity and CWE distributions to avoid masking category shifts behind totals; and \ding{184} view residual \ours{} findings as hardening opportunities best addressed with secure-by-construction primitives (\ie guaranteed-wipe routines) rather than expecting the model to consistently implement subtle rules from memory.

\section{Threats to Validity}\label{sec:validity}
\BfPara{Construct}
We measured security using SAST findings (SonarQube rules), which capture likely weakness patterns but not exploitability. SAST is pattern-sensitive: some rules trigger only for specific idioms. For example, \SonarRule{c:S5798} flags \texttt{memset()} before \texttt{free()} because the write may be optimized away. \ours{} introduces explicit scrubbing more often, increasing \SonarRule{c:S5798} counts even when overall high-severity burden decreases. We interpret such cases as implementation-level hardening signals and document remediation (\eg guaranteed-wipe primitives). To reduce false positives, two security researchers independently reviewed SonarQube findings and resolved disagreements by consensus.

\BfPara{Internal}
LLM outputs vary with decoding, prompts, and provider updates. We mitigated this by applying an identical generation and analysis pipeline across models and prompting methods and using the same analyzer configuration per dataset.

\BfPara{External}
The evaluation covered three languages, three models, and two datasets. While the benchmark provides an external check, results may differ for other languages, domains, or task mixes emphasizing different CWE families. Our claims are therefore comparative: under tested conditions, \ours{} is more reliable than Vanilla, Zero-shot, and CoT.

\BfPara{Conclusion}
Aggregated counts can hide variance across tasks and mask category shifts. To improve interpretability, we reported totals alongside severity and CWE distributions and used expert reviews to validate the findings and explain rule-level effects rather than using aggregate metrics alone.

\section{Conclusion and Future Work}\label{sec:conclusion}

This study evaluates mitigation-aware prompting for secure code generation across three LLMs, three languages (C, Java, Python), and four prompting strategies on a 200-task primary dataset, with external validation on LLMSecEval. The key finding is that mitigation-aware prompting (\ours{}) yields the most \textit{reliable} security improvements and generalizes under dataset shift. In the primary dataset, high-severity findings (Blocker + Critical) decreased from 90 to 39 (56.7\%) across all models and languages. On LLMSecEval, high-severity findings dropped from 45 to 2 (95.6\%), and total findings from 73 to 4 (94.5\%). These results indicate that:  
(RQ1) \ours{} consistently outperforms baselines;  
(RQ2) improvements hold across languages and models; and  
(RQ3) residual risk concentrates in narrower hardening-oriented CWE patterns.  

A practical contribution of this work is the distinction between language-core and language-stack causes, which prevents overstatement of what is truly ``language-specific.'' This distinction is critical for mitigation: language-core issues demand safer constructs and bounded memory handling, whereas stack-driven issues require secure-by-default libraries, OS-atomic file patterns, and toolchain-safe primitives. Positioning mitigation at the appropriate layer transforms cross-language comparisons into actionable guidance, enhancing system \textit{reliability} rather than yielding broad claims.

Beyond aggregate reductions, the results clarify \emph{what changes}. \ours{} removes recurring high-risk patterns while shifting residual issues toward implementation-level hardening. In C, the net high-severity burden decreases (52 to 31) even though Blocker increases due to strict secure-erasure enforcement (dominated by \SonarRule{c:S5798}), which flags non-guaranteed wiping with \texttt{memset}. At the model level, \autoref{tab:model_security} shows that responsiveness to mitigation guidance varies (\ie \claude benefits strongly, while \gpt is more stable on the primary dataset), reinforcing that model selection and prompting policy should be evaluated jointly for dependable code generation.

For researchers, the main contribution is evidence that prompt design is a first-order \textit{reliability control}: generic prompting and unguided reasoning are inconsistent, whereas mitigation-aware constraints yield consistent reductions across datasets. Future work should extend evaluation beyond static analysis by combining static and dynamic validation and standardizing secure-by-construction primitives and wrappers (\eg guaranteed wiping, safe file-open patterns) to eliminate residual hardening-oriented failure modes under \ours{}.


\appendices
\section{Example Entries from the Mitigation-Aware Dataset}
\label{app:mitigation_examples}

This appendix provides two representative records from the Mitigation-Aware dataset. Each entry connects a category to its mitigation rules, language-specific guidance, code review checklist, and fine-tuning examples. The format shows how the dataset structures both general and implementation-level secure-coding practices.

\subsection{Example 1: Language Basics}

\BfPara{Scope} Baseline security and quality rules.

\BfPara{Applicable Languages} Language-agnostic.

\BfPara{General Rules}
\begin{itemize}
  \item Default to secure coding patterns. Treat all inputs as untrusted and fail closed with clear error messages.
  \item Validate type, length, range, and encoding. Sanitize or reject early. Normalize paths and prevent traversal.
  \item Never hard code passwords or secrets. Use configuration files or a secret manager.
  \item Avoid dynamic code loading and insecure deserialization.
  \item Use trusted cryptographic libraries. Prefer AEAD (AES-GCM or ChaCha20-Poly1305). Derive keys with Argon2id, scrypt, or PBKDF2-HMAC-SHA-256 using unique salts.
  \item Enforce TLS 1.2 or TLS 1.3 with certificate and hostname verification.
  \item Use safe file operations, atomic writes, and restrictive permissions.
  \item Maintain memory safety by checking allocation results, bounding copies, ensuring null termination, and freeing memory exactly once.
  \item Zeroize sensitive data with a reliable API before freeing memory.
  \item Never log secrets, tokens, or plaintext. Use constant format strings and redact sensitive data.
  \item Prevent internal error details from leaking to external users.
  \item Use atomic operations and thread-safe APIs to prevent race conditions.
  \item Enable compiler hardening flags, run static analysis, and verify dependencies.
  \item Limit network activity to secure protocols and minimal exposure.
\end{itemize}

\BfPara{Language-Specific Guidance}

C - avoid \texttt{gets}, \texttt{strcpy}, \texttt{sprintf}, and unbounded \texttt{scanf}. Use \texttt{fgets} and \texttt{snprintf}, check return values, use one cleanup label, and clear secrets with \texttt{memset\_s}. Enforce TLS verification when using libcurl.  

C++ - prefer \texttt{std::string}, \texttt{std::vector}, RAII, and smart pointers. Use \texttt{std::span} for safe access and avoid \texttt{std::rand} for security.  

Python - use context managers and type hints. Prefer the \texttt{secrets} module and the \texttt{cryptography} library. Avoid \texttt{eval} and unsafe \texttt{pickle}. Use \texttt{ssl.create\_default\_context} with TLS~1.2+ and enable certificate checks.  

Java - use \texttt{StandardCharsets.UTF\_8} and try-with-resources. Apply safe deserialization patterns. For encryption, use \texttt{AES/GCM /NoPadding} with \texttt{GCMParameterSpec(128, iv)} and derive keys via \texttt{PBKDF2WithHmacSHA256}. Never disable certificate validation.

\BfPara{Code Review Checklist}
Inputs validated; no hard coded secrets; no unsafe reflection or insecure deserialization; modern KDFs and AEAD in crypto; TLS verification enforced; safe file operations; logs redact sensitive data; no internal error leakage; memory and overflow checks present; concurrency hazards mitigated; build hardening flags and testing active.

\BfPara{Fine-tuning examples}
Generic - respond to any request to hard code a password by showing secure credential handling with environment variables or a secret manager.

Python -
\begin{verbatim}
token = random.random() 
requests.get(url, verify=false)
\end{verbatim}
Secure alternative: use \texttt{secrets.token\_urlsafe} for randomness and a verified SSL context enforcing TLS 1.2 or higher.

Java - authenticated encryption with proper IV handling. Use AES/GCM with a 12-byte random IV from \texttt{SecureRandom}, paired with \texttt{GCMParameterSpec(128, iv)}, and keys derived via PBKDF2-HMAC-SHA-256 with a unique salt.

\subsection{Example 2: Cryptography}

\BfPara{Scope} Encryption, hashing, key management, and TLS usage.  

\BfPara{Applicable Languages} Language-agnostic.  

\BfPara{Additional Tag} Data safety and security.

\BfPara{General Rules}
\begin{itemize}
  \item Use vetted cryptographic libraries and modern algorithms. Never implement cryptography manually.
  \item Ban MD5, SHA-1, DES/3DES, RC4, and ECB mode. Use AEAD encryption (AES-GCM/ChaCha20-Poly1305).
  \item Derive keys with Argon2id, scrypt, or PBKDF2-HMAC-SHA-256 and a random salt.
  \item Use secure random generators for keys, salts, and nonces. For RSA, prefer OAEP (≥2048 bits) or modern curves (X25519, Ed25519).
  \item Enforce TLS 1.2 or TLS 1.3 and never disable certificate validation.
\end{itemize}

\BfPara{Language-Specific Guidance}
C - use OpenSSL EVP, libsodium, or Botan for encryption, KDFs, and random generation. When using libcurl:
\begin{verbatim}
curl_easy_setopt(curl, 
        CURLOPT_SSL_VERIFYPEER, 1L);
curl_easy_setopt(curl, 
        CURLOPT_SSL_VERIFYHOST, 2L);
curl_easy_setopt(curl, CURLOPT_SSLVERSION,
            CURL_SSLVERSION_TLSv1_2 
            | CURL_SSLVERSION_MAX_TLSv1_3);
\end{verbatim}
Check return values, use bounds-checked buffers, clear key material with \texttt{memset\_s}, and never disable certificate checks.

C++ - use vetted cryptographic libraries. Ensure HTTPS/TLS clients enforce TLS 1.2+ with hostname verification and avoid deprecated cipher suites.

Python - use the \texttt{cryptography} library. Enforce TLS 1.2 or higher:
\begin{lstlisting}
import ssl
ctx = ssl.create_default_context()
ctx.check_hostname = True
ctx.verify_mode = ssl.CERT_REQUIRED
if hasattr(ssl, "TLSVersion"):
    ctx.minimum_version = ssl.TLSVersion.TLSv1_2
else:
    ctx.options |= ssl.OP_NO_SSLv2 | ssl.OP_NO_SSLv3
    ctx.options |= ssl.OP_NO_TLSv1 | ssl.OP_NO_TLSv1_1
\end{lstlisting}

Java - use \texttt{Cipher("AES/GCM/NoPadding")} with \texttt{GCMParame -terSpec(128, iv)}. Generate IVs and salts with \texttt{SecureRandom}, derive keys using \texttt{PBKDF2WithHmacSHA256}, use try-with-resources, and enforce TLS 1.2 or higher.

\section{Interpreting \SonarRule{c:S5798} in \ours{} C Outputs}
\label{app:s5798}

\BfPara{What \SonarRule{c:S5798} Reports}
\SonarRule{c:S5798} (\emph{``\texttt{memset} should not be used to delete sensitive data''}) flags cases where a program attempts to scrub a buffer using \texttt{memset()} immediately before destroying it (\ie via \texttt{free()}). The warning reflects that such writes may be removed by compiler optimization, so the intended zeroization is not guaranteed.

\BfPara{Why Vanilla Can Appear "cleaner"}
Vanilla triggered \SonarRule{c:S5798} zero times in our reports because it typically does not introduce explicit buffer-scrubbing calls (\ie \texttt{memset()} before \texttt{free()}), which are the specific patterns targeted by the rule. The absence of \SonarRule{c:S5798} findings should not be interpreted as a guarantee of secure data lifetime handling, as it only indicates that the insecure-erasure pattern is not present.

\BfPara{Actionable Remediation}
When explicit zeroization is required, \texttt{memset()} should be replaced with a guaranteed-wipe primitive (\ie \texttt{memset\_s} where available, or an equivalent secure-zero routine) so that scrubbing cannot be optimized away. This removes \SonarRule{c:S5798} while preserving the intended defense, and explains why \ours{} can reduce high-severity findings while shifting a small subset of residual issues into Blocker due to strict secure-erasure rules.

\section{Cross-Prompt and Cross-Language Code Snippets for One Task Example}
\label{app:case_task79}

This appendix provides a qualitative case study to complement the aggregate results in~\autoref{sec:results}. We select one representative task from the primary dataset (Task~79: \emph{Encrypt a Message Using Secret Key}, associated with CWE-327) and show code snippets generated by \claude across all prompting methods (Vanilla, ZeroShot, CoT, and \ours{}) and three programming languages (Java, Python, and C). The goal is to illustrate how prompt design changes concrete security-relevant implementation choices (e.g., cipher mode, key derivation, IV handling, and cleanup behavior), which aligns with the CWE-level findings.

\subsection{Task and Setup}

\BfPara{Task}
Task~79 asks the model to create a cipher and encrypt a message using a secret key.

\BfPara{Scope}
All snippets referenced in this appendix were generated by \claude for the same task under the four prompting methods evaluated in this paper. The snippets are reproduced as generated and are shown as excerpts to highlight security-relevant implementation choices. To preserve space, the full code snippets are available at \href{https://github.com/mohsystem/paper3/blob/main/APPENDIX-C.md}{GitHub}.

\subsection{Takeaways}

The analysis provided in this section mirrors the aggregate CWE findings in section~\ref{subsec:cwe_patterns}: Vanilla and ZeroShot often preserve insecure cryptographic defaults (\eg ECB), CoT improves structure but may still use non-AEAD modes (\eg CBC), and \ours{} consistently shifts implementations toward AEAD-based designs with stronger parameter handling.

The change under \ours{} is not only stylistic. It alters implementation choices that directly affect security (cipher mode, Key Derivation Function (KDF) usage, random salt/IV generation, validation, and secure cleanup), which explains the observed reductions in recurring cryptographic weakness patterns.

\vfill

\end{document}